\documentclass[12pt,dvips]{article}
\usepackage{eurosym}
\usepackage{graphicx}
\usepackage{amsmath}
\usepackage{amssymb}
\usepackage{latexsym}
\usepackage{color}
\usepackage{epstopdf}

\begin{document}

\thispagestyle{empty}

\begin{center}
\vspace{1cm}

{\Large \textbf{Quantum Coherence Versus Non-Classical Correlations in
		Optomechanics}} 

\vspace{1cm}

\textbf{Y. Lahlou}$^{a}${\footnote{%
email: \textsf{youness$_{-}$lahlou@um5.ac.ma}}}, \textbf{M. Amazioug}$^{b,c}${%
\footnote{%
email: \textsf{amazioug@gmail.com}}},
\textbf{J. El Qars}$^{c,d}${\footnote{%
email: \textsf{j.elqars@gmail.com}}},
\textbf{N. Habiballah}$%
^{c,d,e} ${\footnote{%
email: \textsf{n.habiballah@uiz.ac.ma}}}\textbf{M. Daoud}$^{e,f}${\footnote{%
email: \textsf{m$_{-}$daoud@hotmail.com}}} and \textbf{M. Nassik}$^{c}${\footnote{%
email: \textsf{m.nassik@uiz.ac.ma}}},

\vspace{0.5cm}

$^{a}$\textit{LPHE-MS, Department of Physics, Faculty of Sciences, Mohammed V University, Rabat, Morocco}\\[0.5em]
$^{b}$\textit{Ecole Normale Superieure, Mohammed V University, Rabat, Morocco }\\[0.5em]
$^{c}$\textit{EPTHE, Department of Physics, Faculty of Sciences, Ibn Zohr University, Agadir, Morocco}\\[0.5em]
$^{d}$\textit{Faculty of Applied Sciences, Ibn Zohr University, Ait-Melloul, Morocco}\\[0.5em]
$^{e}$\textit{Abdus Salam International Centre for Theoretical Physics, Strada Costiera, 11, 34151 Trieste, Italy}\\[0.5em]
$^{f}$\textit{Department of Physics, Faculty of Sciences, University Ibn Tofail, Kenitra, Morocco}\\[0.5em]

\vspace{2cm} \textbf{Abstract}
\end{center}

Coherence arises from the superposition principle, where it plays a central role in quantum me-chanics. In [Phys.Rev.Lett.114,210401(2015)], it has been shown that the freezing phenomenon of quantum correlations beyond entanglement, is intimately related to the freezing of quantum cohe-rence (QC). In this paper, we compare the behaviour of entanglement and quantum discord with quantum coherence in two di erent subsystems (optical and mechanical). We use respectively the en-tanglement of formation (EoF) and the Gaussian quantum discord (GQD) to quantify entanglement and quantum discord. Under thermal noise and optomechanical coupling e ects, we show that EoF, GQD and QC behave in the same way. Remarkably, when entanglement vanishes, GQD and QC re-main almost una ected by thermal noise, keeping non zero values even for high temperature, which in concordance with [Phys.Rev.Lett.114,210401(2015)]. Also, we nd that the coherence associated with the optical subsystem are more robust{against thermal noise{than those of the mechanical subsystem. Our results con rm that optomechanical cavities constitute a powerful resource of QC. 
\newpage

\section{Introduction}

Quantum coherence (QC) is one of the most important features of quantum physics \cite{1,2}. Such non-classical property is more fundamental than entanglement, where it arises due to the superposition principle \cite{3}. Unlike quantum correlations (e.g., entanglement \cite{4}, quantum discord \cite{5} and quantum steering \cite{6}), QC can appear among multi-partite systems as well as in single-partite systems \cite{7}.

In the last decades, many works have been shown that QC can constitute an important resource for quantum information tasks \cite{8}. We cite for instance, quantum teleportation \cite{9}, quantum error correction \cite{10}, quantum cryptography \cite{11} and quantum dense coding \cite{12}.

In the Ref \cite{7}, it has been established a rigorous framework for quantifying QC in nite dimensional quantum states, while, in the Ref \cite{13} a framework for quantifying QC in Gaussian states has been provided based on the relative entropy \cite{4}.

On the other hand, QC has long been considered as the border between the quantum and classical worlds \cite{14}. Nowadays, it has been shown that entanglement is also an important ingredient for several quantum information tasks \cite{15}. In general, a system of two or more subsystems is quali ed entangled, when non-separable quantum correlations are sharing between the di erent subsystems. In other words, a quantum state describing two or more subsystems is said to be entangled, if cannot be mathematically written as a simple product of the quantum states associated with the di erent subsystems \cite{4}.

Because of their theoretical as well as experimental interest in quantum information processing, many e orts have been accomplished to quantify entanglement in Gaussian states \cite{4}. In this sense, miscellaneous computable entanglement quanti ers for two-mode Gaussian states (TMGS) were propo-sed,e.g., the logarithmic negativity \cite{16}, the Gaussian Renyi-2 entanglement \cite{17}, and the entanglement of formation \cite{18}.

Notice here that such measures cannot capture globally the non-classical feature of the state under investigation \cite{5}. Indeed, it was proven that some separable states (unentangled states) which are reach a certain level of mixture might also present non-zero quantum correlations,i.e., quantumness of correlations \cite{19}. In this context, Gaussian quantum discord has been initially proposed as an approach to quantify quantum correlations mainly in Gaussian separable states \cite{20}. Later, other measure of quantumness of correlations in TMGS were proposed,e.g., the Gaussian geometric discord \cite{21}, the operational Gaussian Discord \cite{22}, the Gaussian Hellinger distance \cite{23}, and the Gaussian Renyi-2 discord \cite{24}.

We emphasize that the concept of quantum discord has aroused great interest as a resource which is more robust than entanglement versus the decoherence phenomenon \cite{5}, o ering exponential speed up of certain computational algorithms \cite{19}.

In this paper, we give a comparative study{under in uence of thermal noise and optomechanical coupling{between quantum coherence, entanglement and quantum discord as three indicators of non-classicality in a double-cavity optomechanical system. For this, we focus our attention into two
bi-mode Gaussian states. The rst(second) one is composed by two mixed mechanical(optical) modes or equivalently to a mechanical(optical) subsystem.
Quantum optomechanics studies{inside an optomechanical cavity{the interactions between mecha-nical degrees of freedom and optical modes via radiation pressure e ect  \cite{24}. With recent progresses in micro-fabrication techniques, it is possible actually to generate and test various impressive quantum e ects using optomechanical setups. Proposals include cooling a mechanical mode near its ground state  \cite{25,26}, generation of quantum superposition \cite{27}, realizing entanglement between mechanical and/or optical modes  \cite{28,29}, quantum state transfer  \cite{30}, creation of massive quantum superpositions or so-called Schrodingers cat states  \cite{31} and gravitational wave detection  \cite{32}.
		
The article is organized as follows. In Sec. 2, we describe our optomechanical model and we derive the Heisenberg-Langevin equations governing the dynamics of the system under study. In Sec. 3, we linearize these equations, while in Sec. 4, we obtain the explicit formula of the covariance matrix describing four-mode Gaussian state. In Sec. 5, we discuss three indicators of non-classicality in two di erent subsystems(mechanical and optical), and we compare them under in uence of thermal noise and optomechanical coupling. Finally, a conclusion closes the paper.

\section{Model \label{sec2}}
\begin{figure}[th]
\centerline{\includegraphics[width=13cm]{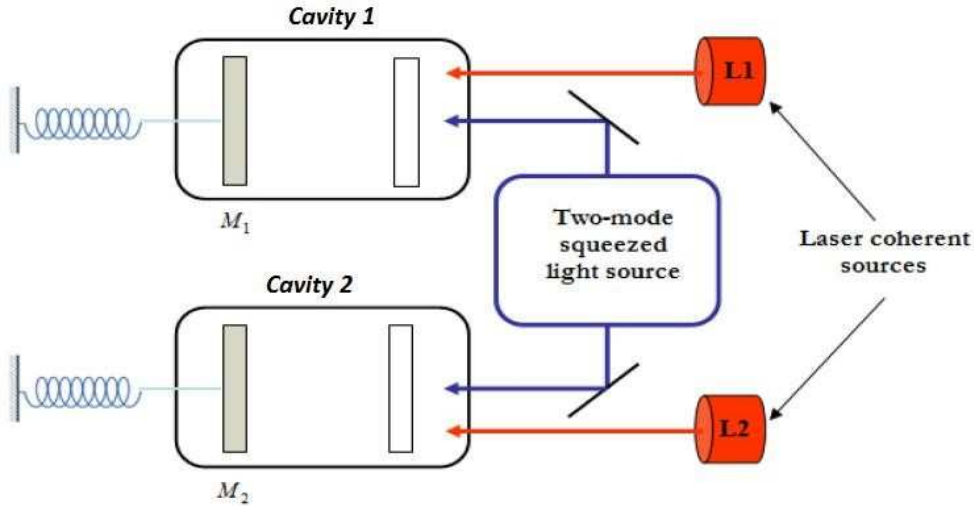}}
\caption{A double-cavity optomechanical system pumped by squeezed light and driven by coherent laser sources.}
\label{Fig.1}
\end{figure}
\noindent In Fig. 1, we consider two Fabry-Perot cavities coupled to a common two-mode squeezed light,
where each cavity is composed by a fixed mirror (partially transmitting) and a movable mirror (labled
${M_P} = 1,2$) perfectly reflecting. The mass and the frequency of the \textit{pth} movable mirror are respectively
${\mu _p}$ and ${\omega _{{M_p}}}$. The \textit{pth} cavity receives a single mode of the input squeezed vacuum light, and pumped
by coherent laser field. Moreover, the \textit{pth} intra-cavity field is coupled to its corresponding movable
mirror via radiation pressure, with coupling rate ${g_p} = \frac{{{\omega _{ap}}}}{{{L_p}}}\sqrt {\frac{\hbar }{{{\mu _p}{\omega _{Mp}}}}} $ \cite{24}, where ${\omega _{ap}}\left( {{L_p}} \right)$ is the frequency(length) of the \textit{pth} cavity.\\ 
In a frame rotating with frequency ${\omega _{{L_p}}}$, the Hamiltonian of the system is given by \cite{33}

\begin{equation}
\mathcal{H}/\hbar  = \sum\limits_{p = 1}^2 {\left[ {{\omega _{{M_p}}}b_p^\dag {b_p} + \left( {{\omega _{ap}} - {\omega _{Lp}}} \right)a_p^\dag {a_p} + {g_p}a_p^\dag {a_p}(b_p^\dag  + {b_p}) + (a_p^\dag {\varepsilon _p}{e^{i{\phi _p}}} + {a_p}{\varepsilon _p}{e^{ - i{\phi _p}}})} \right]} 
\label{equ.1}
\end{equation}

where ${\phi _p}$ and ${\varepsilon _p} = \sqrt {\frac{{2{{\kappa _p}}{\mathcal{P}_p}}}{{\hbar {\omega _{Lp}}}}} $ $(p=1,2)$ are respectively the phase and the strength of the \textit{pth} input
coherent field, and ${\kappa _p}$ is the cavity decay rate. Moreover, ${\mathcal{P}_p}\left( {{\omega _{Lp}}} \right)$ is the drive pump power(the
frequency) of \textit{pth} input laser field. The movable mirrors are considered as quantum harmonic oscillators
with annihilation and creation operators ${b_p}$ and $b_p^ + $, with $\left[ {{b_p},b_p^ + } \right] = 1\quad \left( {p = 1,2} \right)$. On the other hand, ${a_p}$ and $a_p^ + $ are the annihilation and creation operators of the pth cavity mode, with $\left[ {{a_p},a_p^ + } \right] = 1 \quad \left( {p = 1,2} \right)$. Considering equ(\ref{equ.1}), the non-linear quantum Langevin equations describing the dynamics of the
movable mirrors as well as the optical modes are written as \cite{34} :
\begin{equation}
{{\dot b}_p} =  - \left( {i{\omega _{{M_p}}} + \frac{{{\gamma _p}}}{2}} \right){b_p} - i{g_p}a_p^ + {a_p} + \sqrt {{\gamma _p}} b_p^{in}
\end{equation}

\begin{equation}
{{\dot a}_p} =  - \left( {\frac{{{\kappa _p}}}{2} - i{\Delta _p}} \right){a_p} - i{g_p}{a_p}\left( {b_p^ +  + {b_p}} \right) - i{\varepsilon _p}{e^{i{\phi _p}}} + \sqrt {{\kappa _p}} a_p^{in}
\end{equation}
where ${\Delta _p}$ and ${\Delta _p} = {\omega _{{L_p}}} - {\omega _{{a_p}}}$ are respectively the mechanical damping rate and the laser detuning $\left( {p = 1,2} \right)$. $b_p^{in}$ is the \textit{pth} noise operator describing the coupling between the \textit{pth} mechanical mode and its own environment, whereas, $a_p^{in}$ is the squeezed vacuum operator acting on the \textit{pth} optical cavity mode.\\
For a large value of the mechanical quality factor ${\mathcal{O}_p} = {\omega _{{M_p}}}/{\gamma _p} \gg 1$ the mechanical baths can be considered as Markovian \cite{35}. So, assumed this, we have the non-zero correlation relations \cite{35,36}
\begin{equation}
\left\langle {b_p^{in}\left( \omega  \right)b_p^{in + }\left( {\omega '} \right)} \right\rangle  = \left( {{n_{t{h_p}}} + 1} \right)\delta \left( {\omega  + \omega '} \right)
\end{equation}
\begin{equation}
\left\langle {b_p^{in + }\left( \omega  \right)b_p^{in}\left( {\omega '} \right)} \right\rangle  = {n_{t{h_p}}}\delta \left( {\omega  + \omega '} \right)
\end{equation}

where ${n_{t{h_p}}} = {\left[ {\exp \left( {\frac{{\hbar {\omega _{{M_p}}}}}{{{k_B}{T_p}}}} \right) - 1} \right]^{ - 1}}$ is the mean phonons number of the \textit{pth} thermal bath with temperature ${T_p}$. $k_B$ is the Boltzmann constant.
The squeezed vacuum operator $a^{in}_p$ has the following non-zero correlations properties \cite{37}

\begin{equation}
\left\langle {a_p^{in}\left( \omega  \right)a_p^{in + }\left( {\omega '} \right)} \right\rangle  = \left( {N + 1} \right)\delta \left( {\omega  + \omega '} \right),p \in \left\{ {1,2} \right\}
\end{equation}
\begin{equation}
\left\langle {a_p^{in + }\left( \omega  \right)a_p^{in}\left( {\omega '} \right)} \right\rangle  = N\delta \left( {\omega  + \omega '} \right),p \in \left\{ {1,2} \right\}
\end{equation}
\begin{equation}
\left\langle {a_q^{in}\left( \omega  \right)a_p^{in + }\left( {\omega '} \right)} \right\rangle  = M\delta \left( {\omega  + \omega ' - 2{\omega _M}} \right),q \ne p \in \left\{ {1,2} \right\}
\end{equation}
\begin{equation}
\left\langle {a_q^{in + }\left( \omega  \right)a_p^{in}\left( {\omega '} \right)} \right\rangle  = M\delta \left( {\omega  + \omega ' + 2{\omega _M}} \right),q \ne p \in \left\{ {1,2} \right\}
\end{equation}
where $\mathcal{N} = {\sinh ^2}r,\quad \mathcal{M} = \sinh r\cosh r$ with $r$ is the squeezing parameter.
\section{Linearization of quantum Langevin equations}
The non-linear quantum Langevin equations (2)-(3) are in general non solvable analytically. In this way, we use the scheme of linearization given in Ref. \cite{38},i.e., we write each operator as follows
\begin{equation}
{b_p} = {\bar b_p} + \delta {b_p}\quad;\quad{a_p} = {\bar a_p} + \delta {a_p}
\end{equation}
where $\delta {b_p}$ and $\delta {a_p}$ are the operators of fluctuations. ${\bar b_p}$ and ${\bar a_p}$ are respectively the mean values of the operators ${b_p}$ and ${a_p}$. Considering Eqs. (2)-(3) in their steady state, one can obtain :
\begin{equation}
{\bar a_p} = \frac{{ - i{\varepsilon _p}{e^{i{\phi _p}}}}}{{\frac{{{\kappa _p}}}{2} - i{{\Delta '}_p}}}\quad;\quad{\bar a_p} = \frac{{ - i{g_p}{{\left| {{{\bar a}_p}} \right|}^2}}}{{\frac{{{\gamma _p}}}{2} - i{\omega _{{M_p}}}}}
\end{equation}
where ${\Delta '_p} = {\Delta _p} - {g_p}\left( {{{\bar b}_p} + \bar b_p^*} \right)$ is the effective cavity detuning which depends on the displacement of the mirrors due to the radiation pressure force \cite{24}. Replacing (10) in Eqs. (2) and (3), thus we have
\begin{equation}
\delta {{\dot b}_p} =  - \left( {i{\omega _{{M_p}}} + \frac{{{\gamma _p}}}{2}} \right)\delta {b_p} + {\mathcal{G}_p}\left( {\delta {a_p} - \delta a_p^ + } \right) + \sqrt {{\gamma _p}} b_p^{in}
\end{equation}
\begin{equation}
\delta {{\dot a}_p} =  - \left( {\frac{{{\kappa _p}}}{2} - i{{\Delta '}_p}} \right)\delta {a_p} - {\mathcal{G}_p}\left( {\delta b_p^ +  + \delta {b_p}} \right) + \sqrt {{\kappa _p}} a_p^{in}
\end{equation}
where ${{\cal G}_p} = {g_p}\left| {{{\bar a}_p}} \right|$ is the \textit{pth} effective optomechanical coupling \cite{24}. ${\phi _p}$ is an arbitrary phase of the \textit{pth} input laser, which can be chosen to be ${\phi _p} =  - \arctan \left( {\frac{{2{{\Delta '}_p}}}{{{\kappa _p}}}} \right)$ and therefore ${\bar a_p} =  - i\left| {{{\bar a}_p}} \right|$ \cite{39}.
Using the notations $\delta {a_p}\left( t \right) = \delta {\tilde a_p}\left( t \right){e^{i{{\Delta '}_p}t}}$, $\delta {b_p}\left( t \right) = \delta {\tilde b_p}\left( t \right){e^{ - i{\omega _{{M_p}}}t}}$, $\tilde a_p^{in} = {e^{ - i{{\Delta '}_p}t}}a_p^{in}$ and $\tilde b_p^{in} = {e^{i{\omega _{{M_p}}}t}}b_p^{in}$, the equations (12) and (13) became
\begin{equation}
\delta {{\dot \tilde b}_p} =  - \frac{{{\gamma _p}}}{2}\delta {{\tilde b}_p} + {{\cal G}_p}\left( {\delta {{\tilde a}_p}{e^{i\left( {{{\Delta '}_p} + {\omega _{{M_p}}}} \right)t}} - \delta \tilde a_p^ + {e^{ - i\left( {{{\Delta '}_p} - {\omega _{{M_p}}}} \right)t}}} \right) + \sqrt {{\gamma _p}}\tilde b_p^{in}
\end{equation}
\begin{equation}
\delta {{\dot \tilde a}_p} =  - \frac{{{\kappa _p}}}{2}\delta {{\tilde a}_p} - {{\cal G}_p}\left( {\delta \tilde b_p^ + {e^{ - i\left( {{{\Delta '}_p} - {\omega _{{M_p}}}} \right)t}} + \delta {{\tilde b}_p}{e^{ - i\left( {{{\Delta '}_p} + {\omega _{{M_p}}}} \right)t}}} \right) + \sqrt {{\kappa _p}} \tilde a_p^{in}
\end{equation}
Next, using the rotating wave approximation (RWA) \cite{24,30} (i.e. ${\omega _{{M_p}}} \gg {\kappa _p}$ with $p = 1; 2$), the effective cavity detuning is reduced to ${\Delta '_p} \approx {\Delta _p}$, and one can neglect the terms rotating at $ \pm 2{\omega _{{M_p}}}$. In addition, when the cavity is driven at the red sideband (${\Delta '_p} =  - {\omega _{{M_p}}}$ with $p = 1; 2$), the equations (14) and (15) can be written
\begin{equation}
\left( \begin{array}{l}
\delta {{\dot \tilde b}_p}\\
\delta {{\dot \tilde a}_p}
\end{array} \right) = \left( {\begin{array}{*{20}{c}}
	{ - \frac{{{\gamma _p}}}{2}}&{{{\cal G}_p}}\\
	{ - \frac{{{\kappa _p}}}{2}}&{ - {{\cal G}_p}}
	\end{array}} \right)\left( \begin{array}{l}
\delta {{\tilde b}_p}\\
\delta {{\tilde a}_p}
\end{array} \right) + \left( \begin{array}{l}
\sqrt {{\gamma _p}} \tilde b_p^{in}\\
\sqrt {{\kappa _p}} \tilde a_p^{in}
\end{array} \right)
\end{equation}
Finally, from Eq. (16) and using the Fourier transform, one can obtain
\begin{equation}
\delta {{\tilde b}_p}\left( \omega  \right) =  - \frac{{{{\cal G}_p}}}{{b\left( \omega  \right)}}\sqrt {{\gamma _p}} \tilde b_p^{in}\left( \omega  \right) + \frac{{\left( {\frac{{{\gamma _p}}}{2} + i\omega } \right)}}{{b\left( \omega  \right)}}\sqrt {{\kappa _p}} \tilde a_p^{in}\left( \omega  \right)
\end{equation}
\begin{equation}
\delta {{\tilde a}_p}\left( \omega  \right) = \frac{{\left( {\frac{{{\gamma _p}}}{2} + i\omega } \right)}}{{{\xi _p}\left( \omega  \right)}}\sqrt {{\gamma _p}} \tilde b_p^{in}\left( \omega  \right) + \frac{{{{\cal G}_p}}}{{{\xi _p}\left( \omega  \right)}}\sqrt {{\kappa _p}} \tilde a_p^{in}\left( \omega  \right)
\end{equation}
where ${\xi _p}\left( \omega  \right) = {\cal G}_p^2 + \left( {\frac{{{\gamma _p}}}{2} + i\omega } \right)\left( {\frac{{{\kappa _p}}}{2} + i\omega } \right)$
\section{Steady state covariance matrix}
The linear quantum Langevin equations allow us to deduce the covariance matrix (CM) that describes the evolution of steady states of the system \cite{4,41}. Therefore, this allows us to characterize the non-classical behaviour between the various pairwise modes using different quantifiers of quantum correlations.\\
For the sake of simplicity, we consider two identical cavities driven by two identical coherent laser sources, and identical thermal baths. So, ${\mu _{1,2}} = \mu $, ${\omega _{{a_{1,2}}}} = {\omega _a}$, ${\omega _{{M_{1,2}}}} = {\omega _M}$, ${\kappa _{1,2}} = \kappa $ and ${\gamma _{1,2}} = \gamma $ and ${T_1} = {T_2} = T\left( {nt{h_{1,2}} = nth} \right)$, etc. To derive the explicit formula of the CM describing the whole system, we consider the EPR-type quadrature operators $\delta {\tilde X_{{j_p}}}$ and $\delta {\tilde Y_{{j_p}}}\left( {j = m\left( o \right)} \right)$ for the mechanical(optical) subsystem) defined by \cite{39}
\begin{equation}
\delta {\tilde X_{{m_p}}} = \frac{{\delta \tilde b_p^ +  + \delta {{\tilde b}_p}}}{{\sqrt 2 }}\quad, \quad \delta {\tilde Y_{{m_p}}} = \frac{{\delta {{\tilde b}_p} - \delta \tilde b_p^ + }}{{i\sqrt 2 }};\quad p = 1,2
\end{equation}
\begin{equation}
\delta {\tilde X_{{o_p}}} = \frac{{\delta \tilde a_p^ +  + \delta {{\tilde a}_p}}}{{\sqrt 2 }}\quad, \quad\delta {\tilde Y_{{o_p}}} = \frac{{\delta {{\tilde a}_p} - \delta \tilde a_p^ + }}{{i\sqrt 2 }};\quad p = 1,2
\end{equation}
The CM-elements of the system in the steady state, are given by \cite{37}
\begin{equation}
{\mathcal{V}_{ii'}} = \frac{1}{{4{\pi ^2}}}\int_{ - \infty }^{ + \infty } {\int_{ - \infty }^{ + \infty } {d\omega d\omega '{e^{ - i\left( {\omega  + \omega '} \right)t}}} } {\mathcal{V}_{ii'}}\left( {\omega ,\omega '} \right)
\end{equation}
where the frequency-domain correlation function between the elements $i$ and $i'$ of the vector ${U^T}\left( t \right) = \left( {\delta {{\tilde X}_{{m_1}}},\delta {{\tilde Y}_{{m_1}}},\delta {{\tilde X}_{{m_2}}},\delta {{\tilde Y}_{{m_2}}},\delta {{\tilde X}_{{o_1}}},\delta {{\tilde Y}_{{o_1}}},\delta {{\tilde X}_{{o_2}}},\delta {{\tilde Y}_{{o_2}}}} \right)$ are defined by \cite{4}
\begin{equation}
{\mathcal{V} _{ii'}} = \frac{1}{2}\left\langle {\left\{ {{U_i}\left( \omega  \right),{U_{i'}}\left( {\omega '} \right)} \right\}} \right\rangle \quad for \quad i,i' = 1;.....;8
\end{equation}
After some algebra, we finally obtain
\begin{equation}
{\cal V} = \left( {\begin{array}{*{20}{c}}
	{\begin{array}{*{20}{c}}
		{{{\cal V}_1}}&0&{{{\cal V}_{13}}}&0\\
		0&{{{\cal V}_1}}&0&{ - {{\cal V}_{13}}}\\
		{{{\cal V}_{13}}}&0&{{{\cal V}_1}}&0\\
		0&{ - {{\cal V}_{13}}}&0&{{{\cal V}_1}}
		\end{array}}&{\begin{array}{*{20}{c}}
		{{{\cal V}_{15}}}&0&{{{\cal V}_{17}}}&0\\
		0&{{{\cal V}_{15}}}&0&{ - {{\cal V}_{17}}}\\
		{{{\cal V}_{17}}}&0&{{{\cal V}_{15}}}&0\\
		0&{ - {{\cal V}_{17}}}&0&{{{\cal V}_{15}}}
		\end{array}}\\
	{\begin{array}{*{20}{c}}
		{{{\cal V}_{15}}}&0&{{{\cal V}_{17}}}&0\\
		0&{{{\cal V}_{15}}}&0&{ - {{\cal V}_{17}}}\\
		{{{\cal V}_{17}}}&0&{{{\cal V}_{15}}}&0\\
		0&{ - {{\cal V}_{17}}}&0&{{{\cal V}_{15}}}
		\end{array}}&{\begin{array}{*{20}{c}}
		{{{\cal V}_2}}&0&{{{\cal V}_{57}}}&0\\
		0&{{{\cal V}_2}}&0&{ - {{\cal V}_{57}}}\\
		{{{\cal V}_{57}}}&0&{{{\cal V}_2}}&0\\
		0&{ - {{\cal V}_{57}}}&0&{{{\cal V}_2}}
		\end{array}}
	\end{array}} \right)
\end{equation}
The sub-covariance matrix of the mechanical(optical) subsystem labled ${{\cal V}_m}\left( {{{\cal V}_o}} \right)$ can be obtained considering the global covariance matrix given by Eq. (23), where
\begin{equation}
{{\cal V}_m} = \left( {\begin{array}{*{20}{c}}
	{{{\cal V}_1}}&0&{{{\cal V}_{13}}}&0\\
	0&{{{\cal V}_1}}&0&{ - {{\cal V}_{13}}}\\
	{{{\cal V}_{13}}}&0&{{{\cal V}_1}}&0\\
	0&{ - {{\cal V}_{13}}}&0&{{{\cal V}_1}}
	\end{array}} \right) \quad;\quad {{\cal V}_o} = \left( {\begin{array}{*{20}{c}}
	{{{\cal V}_2}}&0&{{{\cal V}_{57}}}&0\\
	0&{{{\cal V}_2}}&0&{ - {{\cal V}_{57}}}\\
	{{{\cal V}_{57}}}&0&{{{\cal V}_2}}&0\\
	0&{ - {{\cal V}_{57}}}&0&{{{\cal V}_2}}
	\end{array}} \right)
\end{equation}
with
\begin{equation}
{{\cal V}_1} = \frac{{\kappa C\cosh \left( {2r} \right) + \left( {1 + 2{n_{th}}} \right)\left( {\kappa  + \gamma  + \gamma C} \right)}}{{2\left( {\kappa  + \gamma } \right)\left( {1 + C} \right)}}\quad ;\quad {{\cal V}_{13}} = \frac{{\kappa C\sinh \left( {2r} \right)}}{{2\left( {\kappa  + \gamma } \right)\left( {1 + C} \right)}}
\end{equation}
\begin{equation}
{{\cal V}_2} = \frac{{\left( {\kappa  + \gamma  + \kappa C} \right)\cosh \left( {2r} \right) + \left( {1 + 2{n_{th}}} \right)\gamma C}}{{2\left( {\kappa  + \gamma } \right)\left( {1 + C} \right)}}\quad;\quad{{\cal V}_{57}} = \frac{{\left( {\kappa  + \gamma  + \kappa C} \right)\sinh \left( {2r} \right)}}{{2\left( {\kappa  + \gamma } \right)\left( {1 + C} \right)}}
\end{equation}
where C is the optomechanical cooperativity given by \cite{24}
\begin{equation}
C = \frac{{4G}}{{\gamma \kappa }} = \frac{{8\omega _a^2}}{{\mu \gamma {\omega _M}{\omega _L}{L^2}}}\frac{P}{{\left[ {{{\left( {\frac{\kappa }{2}} \right)}^2} + \omega _M^2} \right]}}
\end{equation}
The matrices Vm and Vo given by Eq. (24) correspond to two-mode symmetric squeezed thermal states \cite{4}, and thus can be written in the following form
\begin{equation}
{\mathcal{V}_j} = \left( {\begin{array}{*{20}{c}}
	{{s_j}}&0&{{k_j}}&0\\
	0&{{s_j}}&0&{ - {k_j}}\\
	{{k_j}}&0&{{s_j}}&0\\
	0&{ - {k_j}}&0&{{s_j}}
	\end{array}} \right) \equiv \left( {\begin{array}{*{20}{c}}
	{{s_{{j_1}}}}&{{k_{{j_1}}}_{{j_2}}}\\
	{{k_{{j_1}}}_{{j_2}}}&{{s_{{j_2}}}}
	\end{array}} \right)\quad for\quad j \in \left\{ {o,m} \right\}
\end{equation}
where the index ${j_1}({j_2})$ represents the first(second) mechanical mode (for $j=m$) or the first(second) optical mode (for $j=o$). ${S_{{j_1}}} = {S_{{j_2}}} = diag({s_j},{s_j})$ are the sub-matrices describing the first and second modes in the considered subsystem, while the correlations between them are described by the submatrix ${K_{{j_1}{j_2}}} = diag({k_j}, - {k_j})$. For the mechanical subsystem $j \equiv m\quad({s_m} = {{\cal V}_1}\quad and\quad {k_m} = {{\cal V}_{13}})$; while, for the optical one ${\rm{j}} \equiv {\rm{o (}}{{\rm{s}}_o}{\rm{  =  }}{{\rm{\mathcal{V}}}_{\rm{2}}}{\rm{ \quad and\quad  }}{{\rm{k}}_{\rm{o}}}{\rm{  =  }}{{\rm{\mathcal{V}}}_{{\rm{57}}}}{\rm{)}}$.
Notice that using Eq. (28), we can investigate various quantities (,i.e.,quantum coherence, entanglement and quantum discord) witnessing the non-classical behaviour of the mechanical and optical
subsystems.
\section{Quantum coherence vs quantum correlations}}
\subsection{Gaussian quantum coherence}
Coherence arises from the superposition principle, where it is responsible of many important quantum
effects such as entanglement and other types of quantum correlations \cite{1}. In the Ref \cite{7}, it has been
established a rigorous framework for quantifying the coherence of finite dimensional quantum states,
while, in \cite{13} a framework for quantifying coherence of Gaussian states has been provided.
A state $\rho$ (of discrete or continuous variables) is said to be incoherent if it is diagonal when it
expressed in a fixed orthonormal basis \cite{13}. A quantum map is called incoherent operation (ICPTP) if
it is completely positive, trace-preserving, and maps any incoherent states into incoherent states. For
the case of Gaussian states, the necessary conditions that any coherence measure (labled $\mathcal{C}$) should
satisfy are\\
$i)$ $\mathcal{C}\left( \rho  \right) \ge 0$ and $\quad \mathcal{C}\left( \rho  \right) = 0$ iff $\rho  \in \mathcal{I}$, where $\mathcal{I}$ denotes the set of all incoherent states.\\
$ii)$ the measure $\mathcal{C}$ is monotone under all incoherent completely positive and trace-preserving (ICPTP)
maps,i.e., $\mathcal{C}\left( \rho  \right) \ge \mathcal{C}\left( {ICPTP\left( \rho  \right)} \right)$.\\
For any one-mode Gaussian state $\rho$ with the block matrix ${\mathcal{S}_{{j_p}}} = diag\left( {{s_j},{s_j}} \right)$ for $p=1,2$ and $j \in \left\{ {o,m} \right\}$, a coherence measure is defined as
\begin{equation}
C\left( \rho  \right) = \mathop {\inf }\limits_\delta  \left\{ {S\left( {\rho \parallel \delta } \right), \quad \delta \quad is\quad an\quad in\quad coherent \qquad state} \right\}
\end{equation}
where $S\left( {\rho \parallel \delta } \right) = tr\left( {\rho \ln \rho } \right) - tr\left( {\rho \ln \delta } \right)$ ; is the relative entropy, inf runs over all incoherent Gaussian
states. The entropy $S\left( \rho  \right) =  - tr\left( {\rho \ln \rho } \right)$ of $\rho$ and $\inf \left[ {tr\left( {\rho \ln \delta } \right)} \right]$ are given by \cite{13,42}
\begin{equation}
S\left( \rho  \right) = f\left( {{\eta _{{j_p}}}} \right)
\end{equation}
\begin{equation}
\inf \left[ {tr\left( {\rho \ln \delta } \right)} \right] = \left( {{{\bar n}_{{j_p}}} + 1} \right)\ln \left( {{{\bar n}_{{j_p}}} + 1} \right) - {\bar n_{{j_p}}}\ln {\bar n_{{j_p}}}
\end{equation}
where ${n_{{j_p}}} = \sqrt {\det {\mathcal{S}_{{j_p}}}} ,{\bar n_{{j_p}}} = \frac{1}{2}\left( {{s_j} - 1} \right)$ (for $p=1,2$ and $j \in \left\{ {o,m} \right\}$) and the function $f(x)$ is defined by
\begin{equation}
f\left( x \right) = \left( {x + \frac{1}{2}} \right)\ln \left( {x + \frac{1}{2}} \right) - \left( {x - \frac{1}{2}} \right)\ln \left( {x - \frac{1}{2}} \right)
\end{equation}
Similar to the one-mode case, Eqs.(30)-(31) can be generalized into multi-mode Gaussian states. So, for the simple case of symmetric two-mode Gaussian state $\rho(\mathcal{V}_j)$ with covariance matrix $(\mathcal{V}_j)$ (Eq. 28) a coherence measure is defined as \cite{13}
\begin{equation}
{C^j}\left( \rho  \right) =  - \sum\limits_{p = 1}^2 {f\left( {{\eta _{{j_p}}}} \right)}  + \sum\limits_{p = 1}^2 {\left[ {\left( {{{\bar n}_{{j_p}}} + 1} \right)\ln \left( {{{\bar n}_{{j_p}}} + 1} \right) - {{\bar n}_{{j_p}}}\ln {{\bar n}_{{j_p}}}} \right]}\quad for\quad j \in \left\{ {o,m} \right\}
\end{equation}
where $\left\{ {{\eta _{{j_p}}}} \right\}_{p = 1}^2 = \left\{ {{\eta _{j, + }},{\eta _{j, - }}} \right\}$ are the set of symplectic eigenvalues of the CM $\mathcal{V}_j$ (Eq. 28) \cite{43},
and ${\bar n_{{j_p}}} = \frac{1}{2}\left( {{s_j} - 1} \right)$ is determined by the \textit{pth}-mode covariance matrix ${{s_{{j_p}}}}$. Moreover, using explicit expressions of ${\eta _{j, + }},{\eta _{j, - }}$ given by \cite{44}
\begin{equation}
{\eta _{j, \pm }} = \sqrt {\frac{{{\Delta _j} \pm \sqrt {\Delta _j^2 - 4\det {\mathcal{V}_j}} }}{2}}
\end{equation}
coherence of the optical(mechanical) sub-system $\mathcal{C}^o(\mathcal{C}^m)$ can be quantified as \cite{13}
\begin{equation}
{\mathcal{C}^j} =  - f\left( {{\eta _{j, + }}} \right) - f\left( {{\eta _{j, - }}} \right) + 2f\left( {{s_j}} \right)\quad for\quad j \in \left\{ {o,m} \right\}
\end{equation}
where ${\Delta _j} = \det {\mathcal{S}_{{j_1}}} + \det {\mathcal{S}_{{j_2}}} + 2\det {\mathcal{K}_{{j_1}{j_2}}}$ and the function f is defined by Eq. (32).
\subsection{Entanglement of formation}
A convenient and useful way to quantify entanglement in continuous variables systems is by means of the entanglement of formation (EoF) \cite{9}. It quantifies the minimal amount of entanglement, which is needed in order to prepare the state by mixing pure entangled states \cite{18}. It is defined as \cite{18}
\begin{equation}
{E_F}\left( \rho  \right) = \inf \left\{ {\sum {{p_k}E\left( {\left| {{\psi _k}} \right\rangle \left\langle {{\psi _k}} \right|} \right)\mid\rho  = \sum {{p_k}\left| {{\psi _k}} \right\rangle \left\langle {{\psi _k}} \right|} } } \right\}
\end{equation}
where ${E\left( {\left| {{\psi _k}} \right\rangle \left\langle {{\psi _k}} \right|} \right)}$ is the amount of entanglement of the pure bipartite state ${\left| {{\psi _k}} \right\rangle }$.\\
Eq. (36) corresponds to an infimum over all (possibly continuous) convex decompositions of the state into pure states with respective entanglement $E\left( \psi  \right) = S\left( {t{r_B}\left[ {\left| {{\psi _k}} \right\rangle \left\langle {{\psi _k}} \right|} \right]} \right)$, where $ S\left( X \right) =  - tr\left( {X\log X} \right)$ is the von Neumann entropy. In general, the derivation of an explicit expression of the EoF for arbitrary states is not a trivial task even for special quantum states. However, analytical expressions of the EoF have been obtained in a few finite-dimensional cases \cite{4}. We cite for instance, two qubit states \cite{45}, isotropic states \cite{46}, and Werner states ;\cite{47}. For continuous variable states, and particularly for symmetric two-mode Gaussian states, EoF has been evaluated [18, 48]. With symmetric CM $\mathcal{V}_j$ (Eq. 28), the EoF can be written as \cite{18,48}
\begin{equation}
E_F^j = \left\{ \begin{array}{l}
f\left( {\frac{{\tilde \theta _{j, - }^2 + 1/4}}{{2\tilde \theta _{j, - }^2}}} \right)\hspace{1cm}iff\quad{{\tilde \theta }_{j, - }} < 1/2\\
0\hspace{3.1cm}iff\quad {{\tilde \theta }_{j, - }} > 1/2
\end{array} \right.\quad for\quad j \in \left\{ {o,m} \right\}
\end{equation}
where $f(x)$ is defined by Eq. (32), and ${{\tilde \theta }_{j, - }}$ is the minimum symplectic eigenvalue of the partially transposed CM given by \cite{44}
\begin{equation}
{\tilde \theta _{j, - }} = \sqrt {\frac{{{{\tilde \Delta }_j} - \sqrt {\tilde \Delta _j^2 - 4\det {{\cal V}_j}} }}{2}} 
\end{equation}
${\tilde \Delta _j} = \det {\mathcal{S}_{{j_1}}} + \det {\mathcal{S}_{{j_2}}} - 2\det {\mathcal{K}_{{j_1}{j_2}}}.$
\subsection{Gaussian quantum discord}
In the previous subsection, we presented the entanglement of formation as a measure of non-separable quantum correlations those can be captured in the two considered Gaussian subsystems. Here, we present the Gaussian quantum discord as a general measure of quantum correlations that can be nonzero even in separable states (i,e, when EoF is zero). For two-mode Gaussian state describing by the covariance matrix (28), the GQD of the considered bi-mode subsystems is given by \cite{20}
\begin{equation}
{D^j} = f\left( {\sqrt {\det {{\cal V}_j}} } \right) - f\left( {{\eta _{j, + }}} \right) - f\left( {{\eta _{j, - }}} \right) + f\left( {{\Phi _j}} \right)
\end{equation}
where the function f and the simplectic eigenvalues ${\eta _{j, \pm }}$ are respectively given by Eq. (32) and Eq. (34), while, ${\Phi _j}$ is defined by \cite{20}
\begin{equation}
{\Phi _j} = \frac{{\sqrt {\det {S_{{j_1}}}}  + 2\det {S_{{j_1}}} + 2\det {K_{{j_1}{j_2}}}}}{{1 + 2\sqrt {\det {S_{{j_1}}}} }} = \frac{{{s_j} + 2s_j^2 - 2k_j^2}}{{1 + 2{s_j}}}
\end{equation}
where the elements $s_j$ and $k_j$ are defined from Eq. (28).
\newpage
\subsection{Results and Discussion}
Fig. 2 shows that for a fixed value of the squeezing parameter $r$, the entanglement of formation EoF, the Gaussian quantum discord GQD and quantum coherence QC of the two mechanical modes as well as the two optical modes decrease with increasing of the mean thermal phonons number \textit{nth}, meaning that the temperature degrades the quantum feature of the two studied subsystems. In addition, Fig. 2 shows that under thermal noise, both mechanical and optical entanglement decay rapidly to zero
\begin{figure}[th]
	\centerline{\includegraphics[width=0.45\columnwidth,height=5cm]{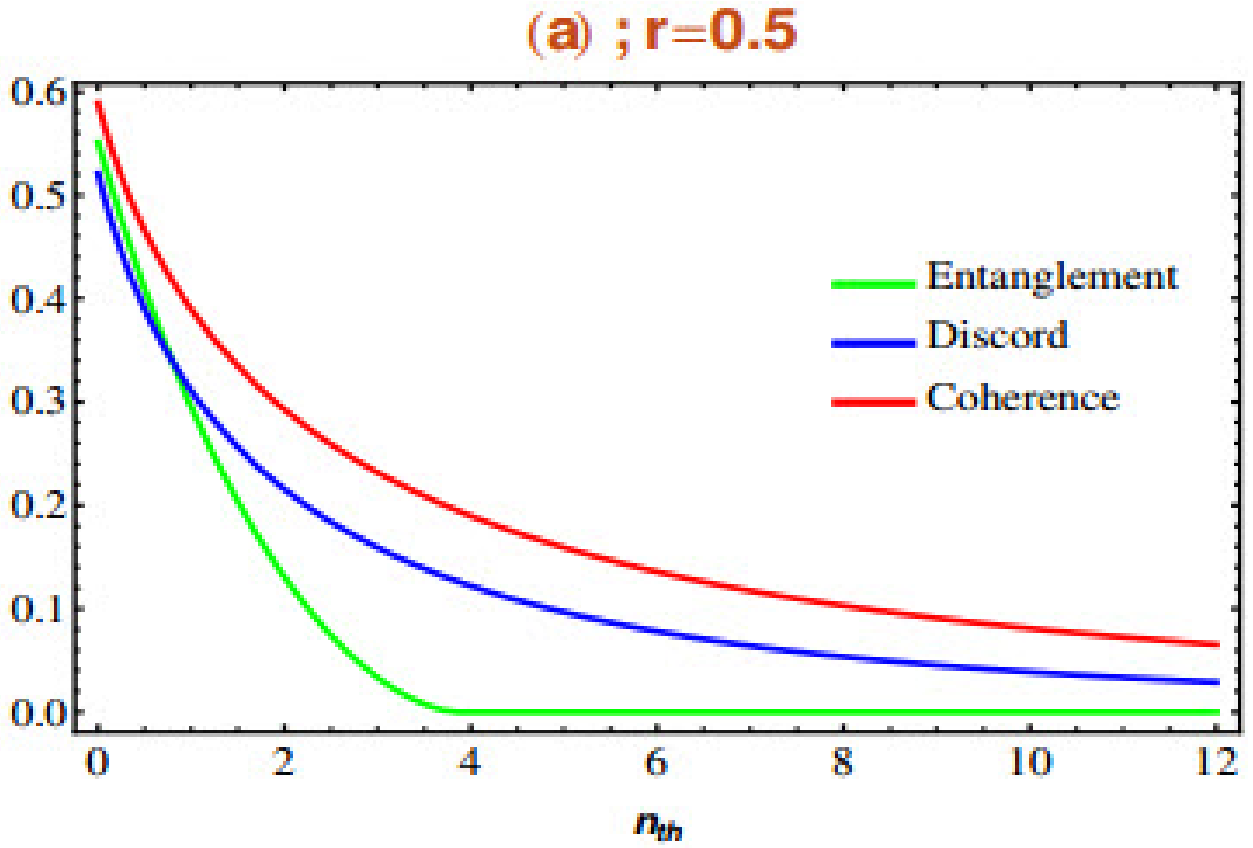}%
		\includegraphics[width=0.45\columnwidth,height=5cm]{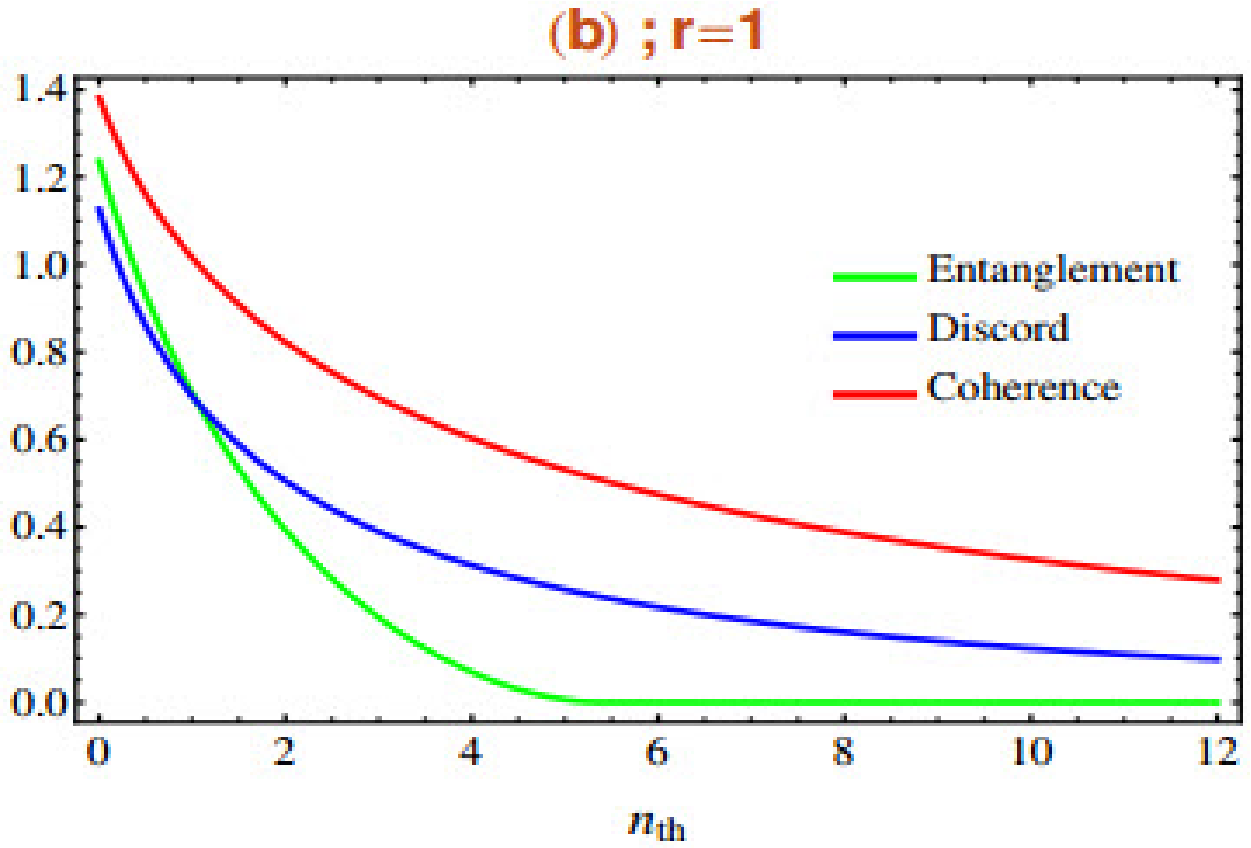}} %
	\centerline{\includegraphics[width=0.45\columnwidth,height=5cm]{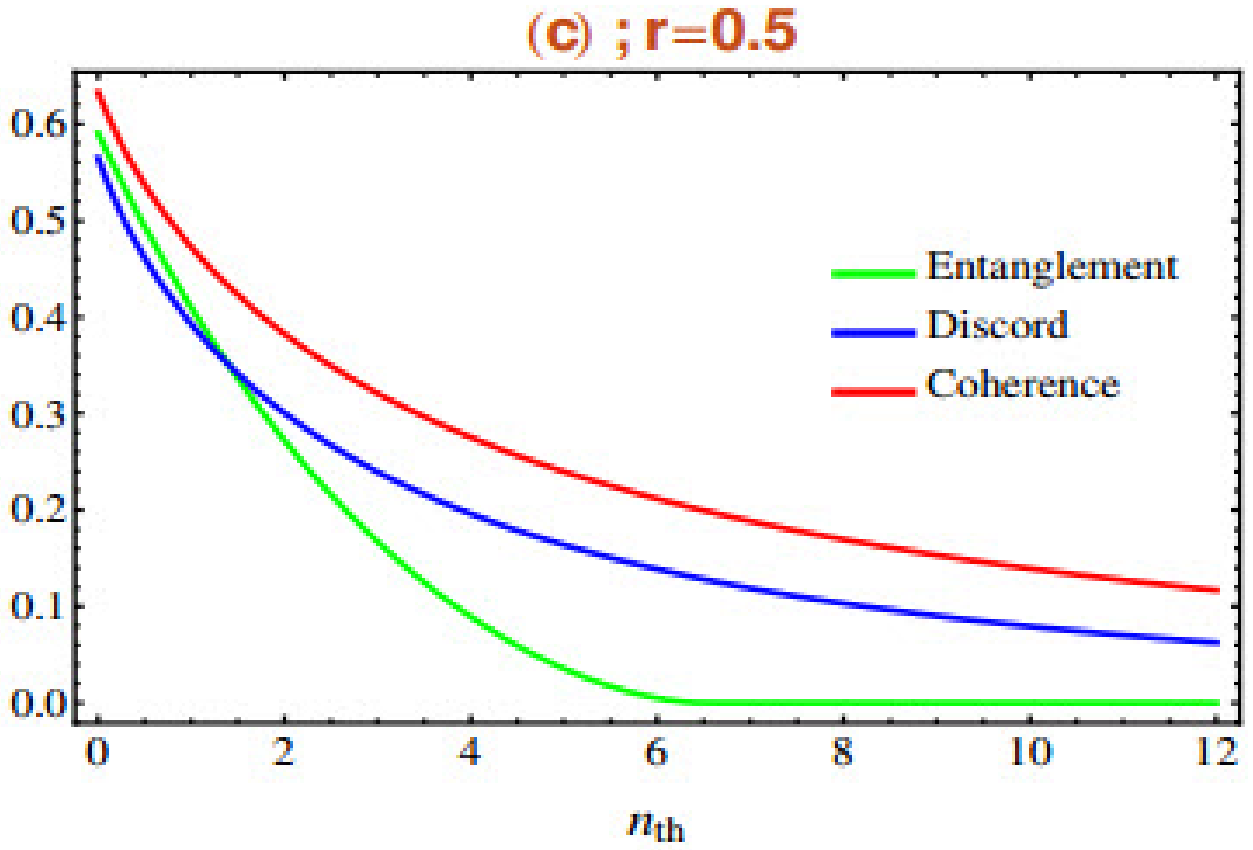}%
		\includegraphics[width=0.45\columnwidth,height=5cm]{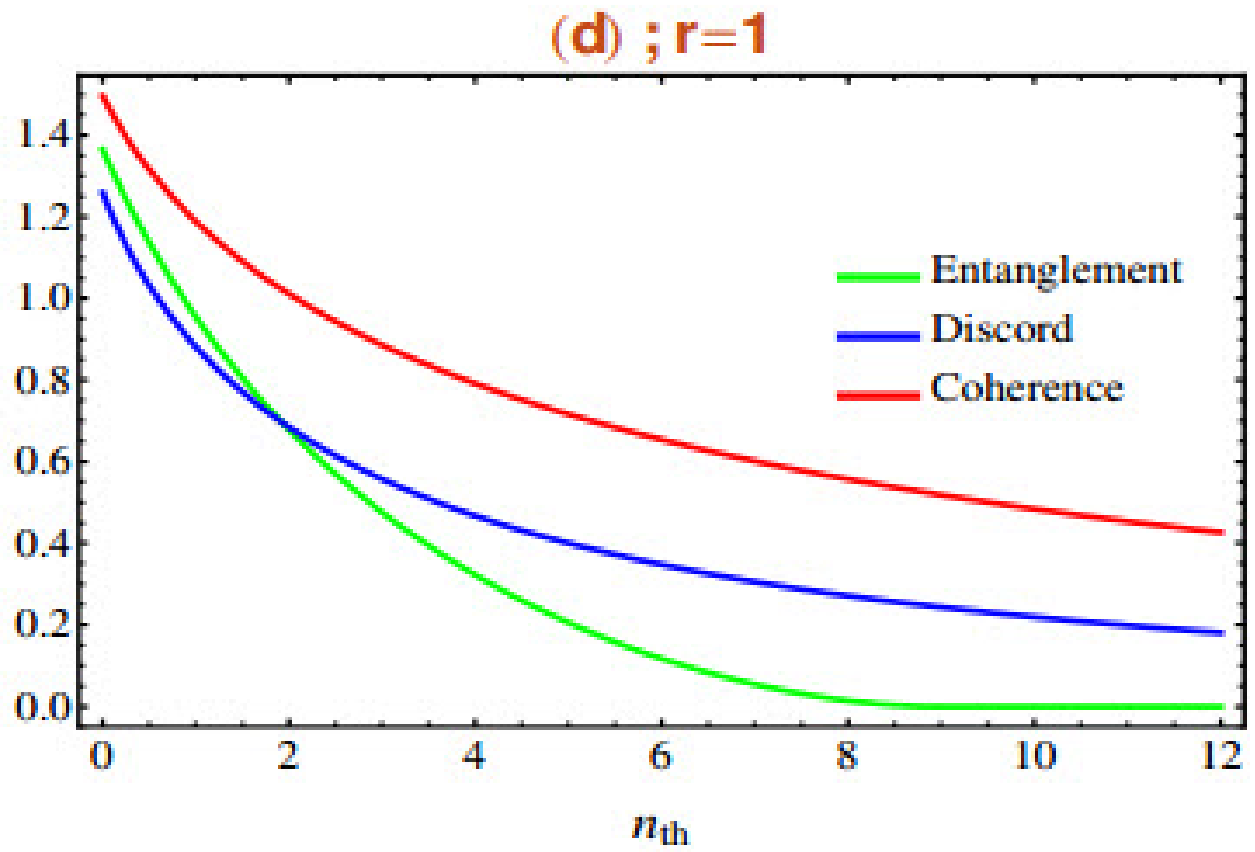}}
	\caption{EoF(green line), GQD(blue line) and QC(red line) versus the thermal phonons number $n_{th}$ for two values of the squeezing $r$. [(a)-(b)] and [(c)-(d)] correspond respectively to the mechanical and optical subsystems. We used $C = 34$ and $\gamma /\kappa = 0.05$.}
	\label{Fig.2}
\end{figure}
comparing with the GQD and QC. In particular, we remark that the mechanical entanglement (Figs.
2(a)-2(b)) is more affected by thermal noise than the optical one (Figs. 2(c)-2(d)). Interestingly, Fig. 2
shows that beyond entanglement (i.e., when EoF=0), GQD and QC undergo a \textit{frozen behaviour} versus
the temperature effect. Indeed, GQD and QC in the two considered subsystems decrease monotonically
when the temperature increases. On the other hand, when entanglement is completely vanishes, the
GQD and QC still persist and remain almost constant{reaching an asymptotic regime{even for high
values of the mean thermal phonons number $n_{th}$ (i.e., high temperature) \cite{49}. More importantly, Fig.
2 reveals a very important behaviour of the GQD and QC under thermal noise. Moreover, in both
mechanical and optical subsystems, the GQD and QC corresponding to the entangled states have a
tendency to decrease drastically under the temperature effect. On the other hand, the GQD and QC
associated with the unentangled states show a robust behaviour,i.e., they decrease asymptotically,
remaining almost constant (\textit{frozen}) even for high values of the mean thermal phonons number $n_{th}$.
We emphasize here that such remarkable cadence corresponding to the freezing phenomenon of GQD
and QC beyond entanglement, is completely in concordance with the physical interpretation given in
\cite{19,49}. Furthermore, comparing Fig. 2(a) with Fig. 2(c); and Fig. 2(b) with Fig. 2(d), we remark
that under thermal effect, quantum coherence measured in the optical subsystem are important than
those measured in the mechanical subsystem. Here, the explanation is that the optical subsystem is
\begin{figure}[th]
	\centerline{\includegraphics[width=0.45\columnwidth,height=5cm]{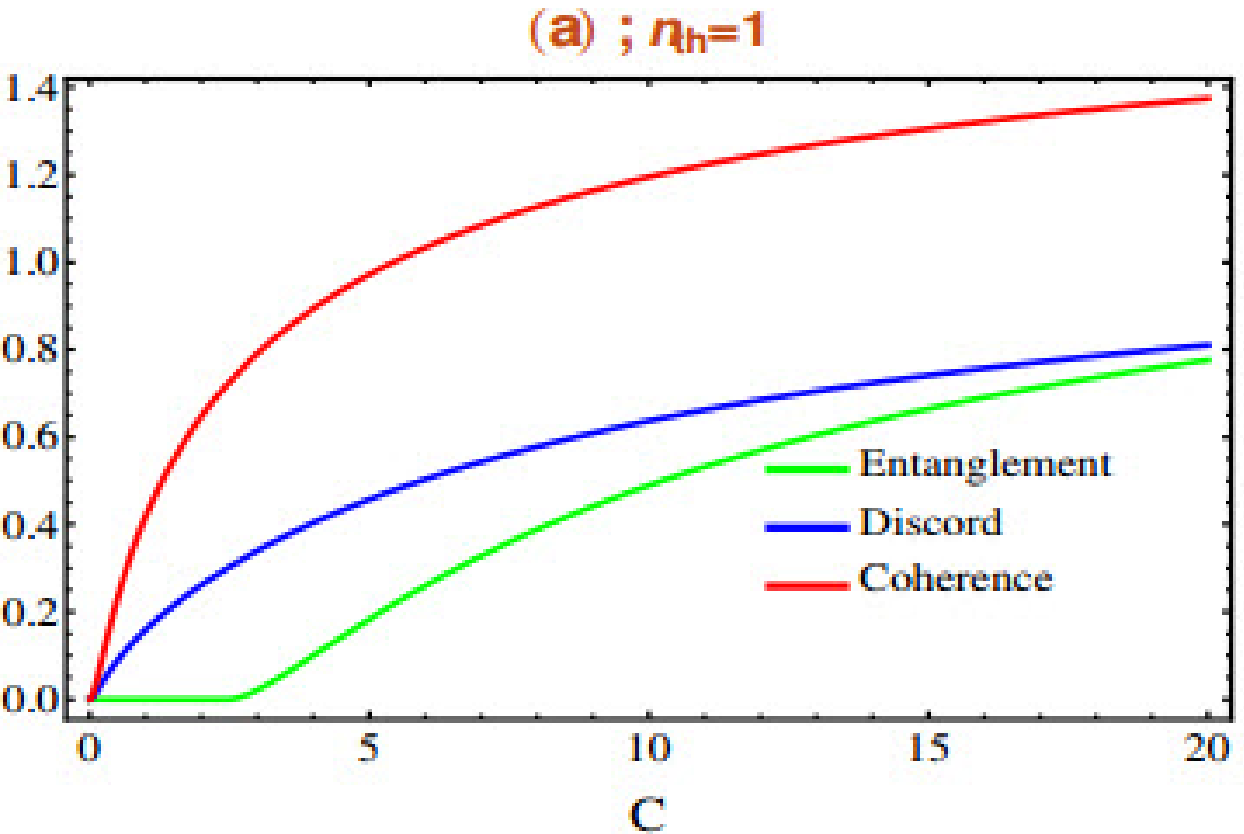}%
		\includegraphics[width=0.45\columnwidth,height=5cm]{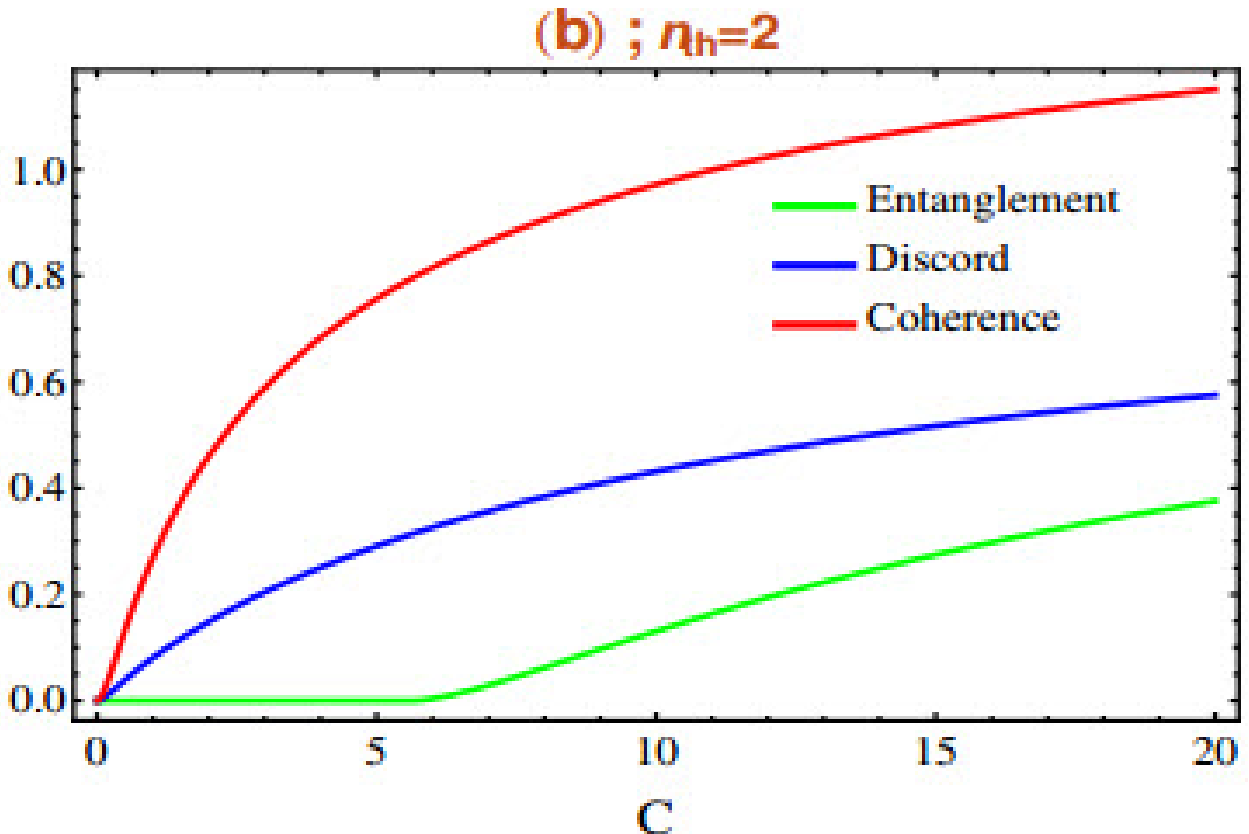}} %
	\centerline{\includegraphics[width=0.45\columnwidth,height=5cm]{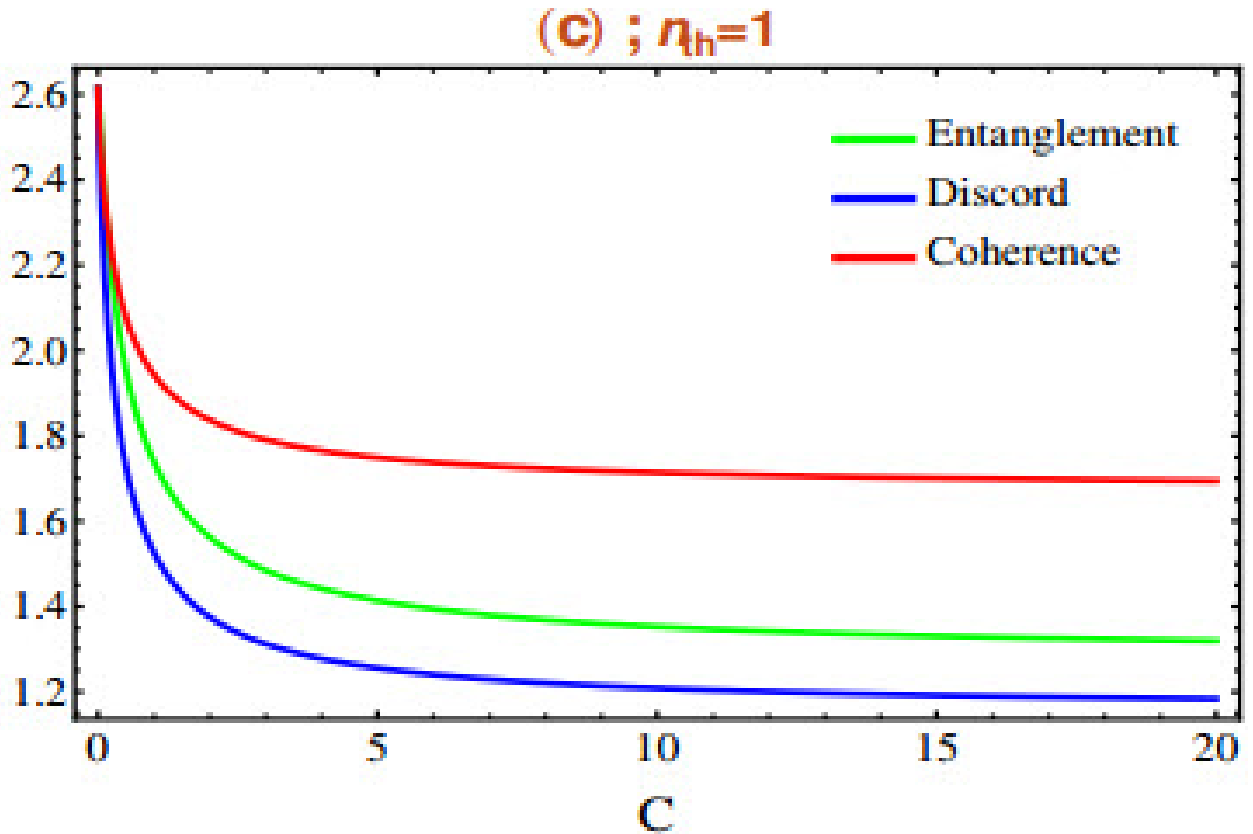}%
		\includegraphics[width=0.45\columnwidth,height=5cm]{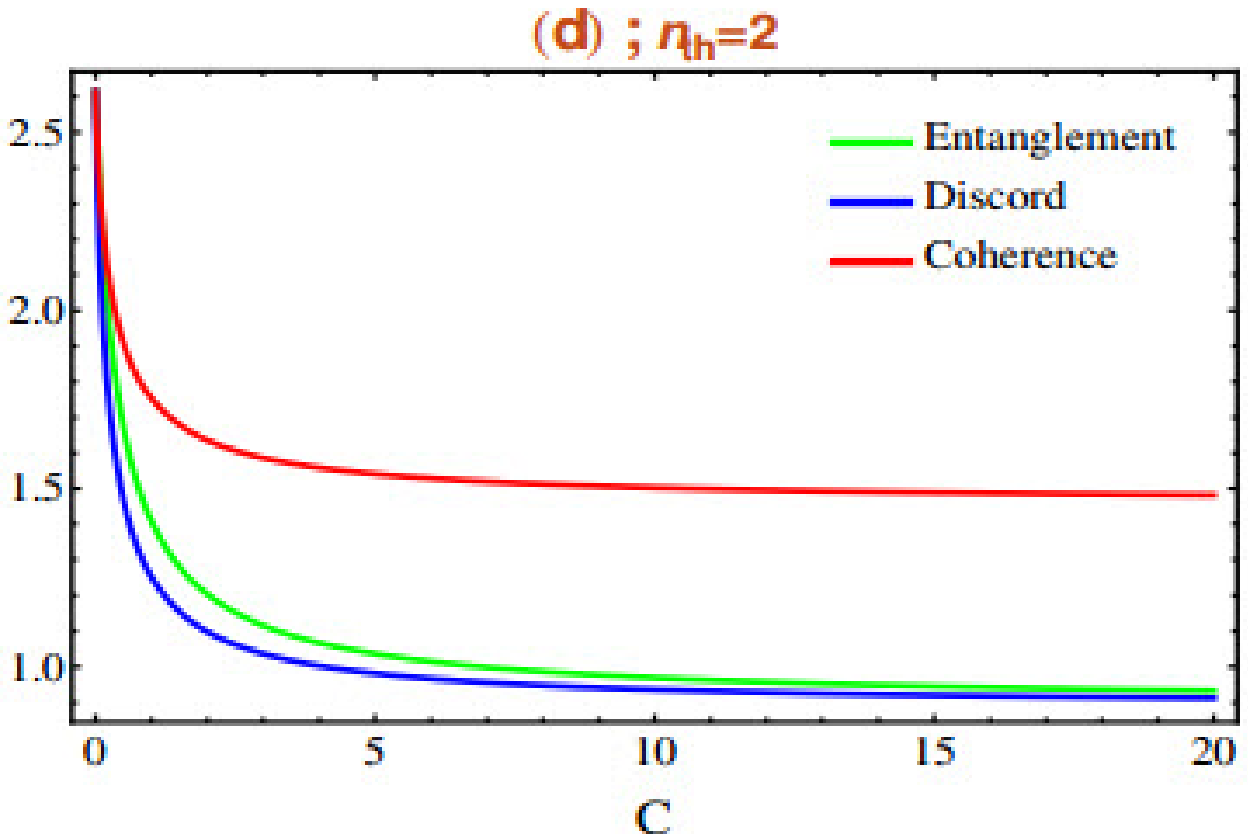}}
	\caption{EoF(green line), GQD(blue line) and QC(red line) versus the the optomechanical cooperativity $C$ for two values of $n_{th}$. [(a)-(b)] and [(c)-(d)] correspond respectively to the mechanical and optical subsystems. We used $r=1.5$ and $\gamma /\kappa = 0.05$.}
	\label{Fig.3}
\end{figure}
influenced by the zero-point fluctuation from the vacuum environment. In contrast, the mechanical
subsystem is influenced by the thermal noise of its bath. So, this justify the fact that the two optical
modes which are in a much more coherent environment than the two mechanical modes, must have a
larger amount of coherence \cite{41}.\\
Next, fixing the squeezing parameter as $r=1.5$, Fig. 3 shows the dependence of the EoF, GQD
and QC of the considered subsystems on the optomechanical cooperativity $C$ for two different values of
the mean phonons number ($n_{th}=1$; $n_{th}=2$). Remarkably, unlike the optical entanglement illustrated in Figs. 3(c)-3(d), Figs. 3(a)-3(b) show that the condition $C\neq 0$ is necessary to create entanglement
between the two mechanical resonators. In addition, Fig. 3 shows that under the same circumstances,
EoF, GQD and GQC of the two optical modes behave in opposite way comparing with those of the
two mechanical modes. Indeed, for a fixed value of the mean thermal phonons number nth, we remark
that when the optomechanical cooperativity C increases, the mechanical EoF, GQD and GQC increase
(Figs. 3(a)-3(b)), whereas, the optical EoF, GQD and GQC decrease (Figs. 3(c)-3(d)). This can be
interpreted as quantum coherence transfer from the optical subsystem to the mechanical subsystem
(i.e., light to matter quantum coherence transfer).\\
Finally, as can be seen from Figs. 2-3, in both mechanical and optical subsystems, GQD and QC
are always non zero even in the regions where EoF is zero. So, on the one hand, non zero GQD in separable states witnesses the presence of \textit{quantumness of correlations} in such states. On the other
hand, non zero QC in these states traduces the non-classical feature of the two considered subsystems
without entanglement.

\section{Conclusions \label{sec4}}

In this work, we have compared the behaviour of entanglement and quantum discord with quantum coherence in two di erent bipartite systems. The rst one is composed by two mixed mechanical modes, while the second is constituted by two mixed optical modes. We used respectively the entanglement of formation EoF and the standard Gaussian quantum discord GQD to quantify entanglement and quantum discord. To quantify quantum coherence QC, we have employed the Gaussian quantum coherence measure proposed in \cite{13}. Essentially, we have interested by the in uence of the thermal noise and the optomechanical coupling on the three studied measures.

In the two considered subsystems, we showed that EoF su ers more than GQD and QC against thermal noise, where it vanishes rapidly with increasing of the temperature. In particular, EoF of the two optical modes is found more robust than the mechanical EoF. This makes optical modes useful for quantum information processing, mainly in dissipative and noisy environments.

Especially, beyond entanglement (i.e., when EoF is zero), it has been found that the behaviour of the GQD follows "uniformly" that of QC. Indeed, both GQD and QC show a resistive behaviour against thermal noise, where they decrease asymptotically with increasing of the mean thermal phonons number $n_{th}$. Moreover, they remain almost constant without vanishing even for high values of $n_{th}$, which corresponds to a \textit{frozen behaviour}.

Interestingly enough, all the obtained results show that the amount of GQD and EoF can never exceed that of QC, meaning that quantum coherence are a resource of non-classical correlations.

Our work constitutes a con rmation{in optomechanics{of the mean result that carried out in the Ref. \cite{49} i.e., beyond entanglement, freezing quantum discord and quantum coherence are two phenomena those can not be separated each to other. In addition, this work may contribute to the understanding of the non-classical behaviour of optomechanical systems in dissipative-noisy en-vironments. This is very interesting for several applications in quantum information processing and communication.

\end{document}